\documentstyle[preprint,revtex]{aps} 
\begin{document}
\draft
\preprint{March 1996}
\begin{title}
A remark on off-diagonal long-range order
\end{title}
\author{C.-A. Piguet, D.F. Wang and C. Gruber}
\begin{instit} 
Institut de Physique Th\'eorique\\
Ecole Polytechnique F\'ed\'erale de Lausanne\\
PHB-Ecublens, CH-1015 Lausanne, Switzerland
\end{instit}
%%%%%%%%%%%%%%%%%%%%%%%%%%%%%%%%%%%%%%%%%%%%%%%%%%%%%%%%%%%%%%%%%%%%%%%
\begin{abstract}
In this work, we study some general property of a strongly
correlated electron system defined on a lattice. Assuming
that the lattice system exhibits off-diagonal long-range order,
we show rigorously that this assumption would lead to 
Meissner effect. 
This generalizes previous results of continuous case to a
lattice electron system.  
\end{abstract}
\pacs{}
\narrowtext

It was well known that both the superfluid and the superconductivity
are characterized by existence of off-diagonal long-range orders. 
Pernose and Onsager pointed out that for interacting boson
systems the Bose-Einstein condensation's mathematical statement is 
nonvanishing off-diagonal long-range order in the reduced density
matrix $\rho_1$ \cite{penrose}. The interacting fermion systems were investigated
by C. N. Yang, and a detailed study and generalization of ODLRO 
were carried out \cite{yang1}. It was shown that the superconductivity in
interacting fermionic systems is characterized by non-vanishing 
off-diagonal long-range order in the reduced density matrix 
$\rho_2$ \cite{yang1}.  

Two basic features of superconductivity are the Meissner
effect and the flux quantization \cite{byers}. If an interacting fermionic 
system has off-diagonal long-range order, one should be able to 
mathematically derive the Meissner effect and the flux quantization.
In fact, there was some recent proof of Meissner
effect as a consequence of existence of ODLRO \cite{sewell}. Later, the proof was extended
to prove the flux quantization as a result of 
non-vanishing ODLRO \cite{nieh}, and to a case of the most general magnetic field
distribution \cite{au}. The considerations were all concerned with interacting
continuous fermion system mathematically described by Schr\"odinger
operator. One would expect a similar proof should be true for 
a lattice fermion system. 

Recently, there have been
intense studies on the possibilities of superconductivities of lattice 
electron systems induced by electron-electron correlations, because of 
the discovery of high temperature superconductivity. 
For the Hubbard model, a notable and the simplest model describing
interacting lattice electron system, some $\eta$ pairing 
states were constructed by Yang in any dimensions \cite{yang2}. 
These states have off-diagonal long-range order, and electron systems in
these states  
should be superconducting (unfortunately, these states
are not the ground states of the Hubbard model). In the following, 
we show that given existence of ODLRO for a lattice electron 
system, one can derive  
Meissner effect. This generalizes the previous results
of continuous quantum systems to lattice electron systems. 

Let us consider a system of interacting electrons in a uniform magnetic field
on a two dimensional square lattice defined by the two basis vectors
$e_{x}$ and $e_{y}$. The hamiltonian for this system is
\begin{equation}
H=t_{x}\sum_{i,\sigma}(e^{i\theta^{x}_{i}}c_{i+e_{x},\sigma}^{+}
c_{i,\sigma}+h.c)+t_{y}\sum_{i,\sigma}(e^{i\theta^{y}_{i}}c_{i+e_{y},\sigma}^{+}
c_{i,\sigma}+h.c)+V(\{n_{i,\sigma}\})
\end{equation}
where $c_{i,\sigma}^{+}$ and $c_{i,\sigma}$ are the creation and annihilation
fermionic operators for an electron at site $i$ with spin $\sigma$, $t_{x}$ and $t_{y}$ are the 
real hopping amplitudes along the $x$ and the $y$ directions respectively. The potential $V$
is any real interaction of the electrons, depending only on the numbers of electrons 
$n_{i,\sigma}=c_{i,\sigma}^{+}c_{i,\sigma}$ 
and the phases $\theta^{x}_{i}$ and $\theta^{y}_{i}$ are such that
\begin{equation}
\theta^{x}_{i}+\theta^{y}_{i+e_{x}}-\theta^{x}_{i+e_{y}}-\theta^{y}_{i}=\frac{e}{\hbar
c}\phi,
\end{equation}
with $\phi$ the magnetic flux through each plaquette of the lattice. 
In the following, we shall demonstrate that assuming existence of 
ODLRO in this lattice system, the only allowed magnetic field is zero.  

The hamiltonian $H$ is not invariant under usual translations,
but it is well known that there exist magnetic translation operators under which
the hamiltonian is invariant both for the continuous system \cite{zak} and for the lattice
system. For the lattice system the two magnetic translation
operators $T_{x}$ and $T_{y}$ are defined by the following unitary
transformations of the electronic operators
\begin{equation}
\begin{array}{lll}
T_{x}: & \hspace{1cm} & c^{+}_{i,\sigma}\rightarrow e^{i\chi^{x}_{i}}c^{+}_{i+e_{x},\sigma}\\
T_{y}: & \hspace{1cm} & c^{+}_{i,\sigma}\rightarrow e^{i\chi^{y}_{i}}c^{+}_{i+e_{y},\sigma}
\label{trans}
\end{array}
\end{equation}
where the phases $\chi^{x}_{i}$ and $\chi^{y}_{i}$ are given by
\begin{eqnarray}
\chi^{x}_{i}&=&\theta^{x}_{i}+i_{y}\phi\nonumber\\
\chi^{y}_{i}&=&\theta^{y}_{i}-i_{x}\phi.
\end{eqnarray}
where $i=(i_{x},i_{y})$. Moreover, they satisfy the following relation  
\begin{equation}
\chi^{x}_{i}+\chi^{y}_{i+e_{x}}-\chi^{x}_{i+e_{y}}-\chi^{y}_{i}=-\frac{e}{\hbar
c}\phi.
\label{chi}
\end{equation}
It is easy to prove that $T_{x}$ and $T_{y}$ commute with $H$ but not with each other
since Eq.(\ref{chi}) gives $e^{i\phi}T_{y}T_{x}=T_{x}T_{y}$.

On a lattice, the ODLRO is defined by the requirement that 
\begin{equation}
C_{i,j}=<c_{i,\uparrow}^{+}c_{i,\downarrow}^{+
}c_{j,\uparrow}c_{j,\downarrow}>
\neq 0\;\;\;{\rm when}\;\;\;\mid i-j\mid\rightarrow\infty
\end{equation}
where the brackets designate the canonical average
\begin{eqnarray}
&&<A>=Tr(\rho A)\nonumber\\
&&\rho=\frac{e^{-\beta H}}{Tre^{-\beta H}}  
\end{eqnarray}
with $\beta$ the inverse of the temperature and the trace is performed over the
Hilbert space with fixed $N$ number of electrons. $C_{i,j}$ can be
seen as matrix elements of a matrix $C$ and thus, we can make a spectral decomposition as
\begin{equation}
C_{i,j}=\sum_{k}\alpha_{k}\psi_{k}(i)^{*}\psi_{k}(j)
\end{equation}
where $\alpha_{k}$ and $\psi_{k}$ are the eigenvalues and eigenvectors of
$C_{i,j}$. If we have ODLRO, then there exist an eigenvalue (for example
$\alpha_{0}$) of order $N$ \cite{yang1}:
\begin{equation}
C_{i,j}\rightarrow \phi(i)^{*}\phi(j)
\;\;\;{\rm when}\;\;\;\mid i-j\mid\rightarrow\infty
\end{equation}
where $\phi(i)=\sqrt{\alpha_{0}}\psi_{0}(i)$.

The magnetic translations operators allow us to express the function $\phi(i)$
at different points. Let use choose a basis $\{\mid n>\}$ of the subspace with
$N$ electrons. Since $T_{x}$ is a unitary operators, $\{\mid T_{x}^{-1}n>\}$ is still
a basis. We can thus express $C_{i,j}$ in these two different basis
\begin{eqnarray*}
C_{i,j}&=&\sum_{n}<n\mid \rho c_{i,\uparrow}^{+}c_{i,\downarrow}^{+
}c_{j,\uparrow}^{+}c_{j,\downarrow}^{+}\mid n>=
\sum_{n}<T_{x}^{-1}n\mid \rho c_{i,\uparrow}^{+}c_{i,\downarrow}^{+
}c_{j,\uparrow}^{+}c_{j,\downarrow}^{+}\mid T_{x}^{-1}n>\\
&=&\sum_{n}<n\mid \rho T_{x}c_{i,\uparrow}^{+}c_{i,\downarrow}^{+
}c_{j,\uparrow}^{+}c_{j,\downarrow}^{+}T_{x}^{-1}\mid n>=
e^{2i(\chi^{x}_{i}-\chi^{x}_{j})}C_{i+e_{x},j+e_{x}}
\end{eqnarray*}
where we have used the fact that $H$ commute with $\rho$ and the definition
(\ref{trans}) of the magnetic translation operators. In the ODLRO limit, we get
\begin{equation}
\phi^{*}(i)\phi(j)=e^{2i(\chi^{x}_{i}-\chi^{x}_{j})}\phi^{*}(i+e_{x})\phi(j+e_{x}),
\end{equation}
and thus
\begin{equation}
\phi(i)=e^{i\lambda_{x}}e^{2i\chi^{x}_{i}}\phi(i+e_{x})
\label{x}
\end{equation}
where $\lambda_{x}$ is a real number independent of the site $i$.
On the other hand, if we use the basis
$\{\mid T_{y}^{-1}n>\}$ we have
\begin{equation}
\phi(i)=e^{i\lambda_{y}}e^{2i\chi^{y}_{i}}\phi(i+e_{y})
\label{y}
\end{equation}
with $\lambda_{y}$ a fixed real number.
With Eqs.(\ref{x}) and (\ref{y}), we can go along the path
$i,i+e_{x},i+e_{x}+e_{y},i+e_{y},i$ to find a condition on the magnetic flux
enclosed by this path:
\begin{eqnarray}
\phi(i)&=&e^{i\lambda_{x}}e^{2i\chi^{x}_{i}}\phi(i+e_{x})\nonumber\\
&=&e^{i\lambda_{x}}e^{2i\chi^{x}_{i}}e^{i\lambda_{y}}e^{2i\chi^{y}_{i+e_{x}}}
\phi(i+e_{x}+e_{y})\nonumber\\
&=&e^{2i\chi^{x}_{i}}e^{i\lambda_{y}}e^{2i\chi^{y}_{i+e_{x}}}
e^{-2i\chi^{x}_{i+e_{x}}}\phi_{i+e_{y}}\nonumber\\
&=&e^{2i\chi^{x}_{i}}e^{2i\chi^{y}_{i+e_{x}}}e^{-2i\chi^{x}_{i+e_{x}}}e^{-2i\chi^{y}_{i}}
\phi(i)\nonumber\\
&=&e^{\frac{2ie}{\hbar c}\phi}\phi(i)
\end{eqnarray}
where we have used Eq.(\ref{chi}) for the last equality. We have found the
following constraint on the flux
\begin{equation}
e^{\frac{2ie}{\hbar c}\phi}=1
\end{equation}
and thus
\begin{equation}
\phi=n\frac{hc}{2e}
\end{equation}
with $n$ an integer. If we take $n$ to be 1, we get a flux which corresponds to a
magnetic field of the order of $10^9 G$, since the area of each 
plaquette is of the order $\AA^2$. This value of magnetic field is impossible practically. 
Then, the only possible choice is $n=0$, which corresponds to zero magnetic field.
This is precisely the Meissner effect.

This result may apply to many interesting electron systems. 
Recently, there have been considerable activities of constructing 
correlated electron Hamiltonians which have the $\eta$ pairing-type states 
as their ground states \cite{korepin1,shen}. One example is 
\begin{equation}
|G>=\sum_{\{x\}} \prod_{i=1}^{N_e} c_{x_i\uparrow}^\dagger c_{x_i\downarrow}^\dagger |0>.
\end{equation}
This wavefunction has non-zero off-diagonal long-range order. 
A lattice system with a ground state given above would not be able to support
nonzero magnetic field. 

In conclusion, we have proved that for a lattice electron system,
the assumption of existence of ODLRO in the ground state (or at finite temperature)
would naturally lead to the Meissner effect. This remark generalizes the previous
consideration of continuous electron system described by Schr\"odinger
operator to a correlated lattice electron system. 
One would also expect that for a lattice electron
system, the existence of ODLRO would also lead to flux quantization.
It would be very interesting to provide a clear mathematical
proof of flux quantization, as a
consequence of existence of ODLRO in a lattice electron system.

This work was supported by the Swiss National Science Foundation. 

%%%%%%%%%%%%%%%%%%%%%%%%%%%%%%%%%%%%%%%%%%%%%%%%%%%%%%%%%%%%%%%%%%%%%%%%%%%%

\end{document}